\documentclass[12pt]{article}
\pdfoutput=1
\usepackage[paper=letterpaper,margin=1.0in]{geometry}
\usepackage{amsfonts,amsmath,amssymb}
\usepackage{graphicx}
\usepackage{cite}
\usepackage{hyperref}

\newcommand{\be}{\begin{equation}}
\newcommand{\ee}{\end{equation}}
\newcommand{\bea}{\begin{eqnarray}}
\newcommand{\eea}{\end{eqnarray}}
\newcommand{\ba}{\begin{array}}
\newcommand{\ea}{\end{array}}
\newcommand{\ben}{\begin{enumerate}}
\newcommand{\een}{\end{enumerate}}
\newcommand{\bi}{\begin{itemize}}
\newcommand{\ei}{\end{itemize}}
\newcommand{\bc}{\begin{center}}
\newcommand{\ec}{\end{center}}
\newcommand{\bfig}{\begin{figure}}
\newcommand{\efig}{\end{figure}}
\newcommand{\bq}{\begin{quotation}}
\newcommand{\eq}{\end{quotation}}
\newcommand{\bt}{\begin{table}}
\newcommand{\et}{\end{table}}
\newcommand{\btab}{\begin{tabular}}
\newcommand{\etab}{\end{tabular}}
\newcommand{\bs}{\begin{slide}}
\newcommand{\es}{\end{slide}}
\newcommand{\nn}{\nonumber}
\newcommand{\eref}[1]{(\ref{#1})}

\newcommand{\comment}[1]{}

\newcommand{\IC}{\mathbb{C}}
\newcommand{\IR}{\mathbb{R}}
\newcommand{\mQ}{\mathbb{Q}}

\newcommand{\mZ}{\mathbb{Z}}
\newcommand{\mT}{\mathbb{T}}

\newcommand{\cN}{{\cal N}}
\newcommand{\cM}{{\cal M}}
\newcommand{\cW}{{\cal W}}

\begin{document}

\vspace*{1.2cm}

\bc

{\Huge \bf Brane Geometry and Dimer Models}

\vspace{1cm}

{\large \bf Yang-Hui He}$^1$\footnote{yang-hui.he.1@city.ac.uk},
{\large \bf Vishnu Jejjala}$^2$\footnote{vishnu@neo.phys.wits.ac.za},
{\large \bf Diego Rodriguez-Gomez}$^3$\footnote{drodrigu@physics.technion.ac.il}

\vspace*{.5cm}

{${}^{1}$ Department of Mathematics, City University, London,\\
Northampton Square, London EC1V 0HB, UK;\\
School of Physics, NanKai University, Tianjin, 300071, P.R.~China;\\
Merton College, University of Oxford, OX1 4JD, UK\\
}
\vspace*{.2cm}
{${}^{2}$ NITheP, School of Physics, and Centre for Theoretical Physics,\\
University of the Witwatersrand, Johannesburg, WITS 2050, South Africa\\
}
\vspace*{.2cm}
{${}^{3}$ Department of Physics, Technion, Haifa, 3200, Israel\\}

\ec

\vspace*{.2cm}
\centerline{\textbf{Abstract}}
\bigskip
The field content and interactions of almost all known gauge theories in AdS$_5$/CFT$_4$ can be expressed in terms of dimer models or bipartite graphs drawn on a torus.
Associated with the fundamental cell is a complex structure parameter $\tau_R$.
Based on the brane realization of these theories, we can specify a special Lagrangian (SLag) torus fibration that is the natural candidate to be identified as the torus on which the dimer lives.
Using the metrics known in the literature, we compute the complex structure $\tau_G$ of this torus.
For the theories on ${\mathbb C}^3$ and the conifold and for orbifolds thereof $\tau_R = \tau_G$.
However, for more complicated examples, we show that the two complex structures cannot be equal and yet, remarkably, differ only by a few percent.
We leave the explanation for this extraordinary proximity as an open challenge.

\newpage
\tableofcontents

\setcounter{footnote}{0}

\section{Introduction}
\label{sec:intro}
The story of D$3$-branes probing conical Calabi--Yau threefolds (CY$_3$) giving rise to examples of AdS$_5$/CFT$_4$ duality is a theme of central importance to modern physics during the last two decades.
The supersymmetric gauge theory lives on the worldvolume of the D$3$-brane while the gravitational description is obtained by considering the background sourced by $N$ D$3$-branes placed at the tip of the CY$_3$.
The Calabi--Yau manifolds under consideration are cones over a five-dimensional Sasaki--Einstein base ${\cal B}$, and the tip of the cone is what the D$3$-branes probe.
The near-brane region of the geometry is AdS$_5\times {\cal B}$.
This implies that the gauge theory on the worldvolume of the branes is a four-dimensional superconformal field theory generically with $\mathcal{N}=1$ supersymmetry.

To date, almost all known explicit pairs of AdS/CFT belong to a particular subclass of non-compact Calabi--Yau manifolds, the so-called {\em toric} manifolds, of which infinite families have been constructed.
The toric description facilitates the algebraic geometry, the differential geometry, as well as the physics: the geometry is encoded entirely into the combinatorics of certain lattice polytopes and classes of explicit metrics have been constructed; so too can the worldvolume physics be succinctly described in terms of a two-dimensional linear sigma model.
We will thus focus on toric Calabi--Yau threefolds.

Even though various techniques for constructing the dual worldvolume field theory given a toric diagram have been developed using the tool of D-brane partial resolution of singularities since the early days \cite{Douglas:1997de,Beasley:1999uz,Feng:2000mi}, it was not until \cite{Hanany:2005ve,Franco:2005rj,Feng:2005gw} that it was realized that the most powerful way of understanding AdS$_5$/CFT$_4$ for toric Calabi--Yau threefolds, is through {\em dimer models}, or, equivalently, brane tilings.

Using dimer models, the gauge theory of interest can be neatly encoded in a bipartite graph drawn on a torus.
This graph expresses the complete information about both the field content and the interactions of the theory: the gauge groups are represented by polygonal faces in the graph, the fields by edges, and the superpotential terms by vertices, which are colored either black and white.
Furthermore, it is possible to encode dynamical data such as the scaling dimensions of the fields at the conformal fixed point in the infrared in terms of the angles between the edges on the bipartite graph, which, in turn, fixes a particular shape for the fundamental cell on the torus.
Quite surprisingly, this shape for the torus --- encoded by the complex structure of the unit cell --- is, upon the obvious action of $SL(2,\,\mathbb{Z})$, an invariant:
all the toric (Seiberg dual) phases of the theory have the same complex structure \cite{Hanany:2011bs}.
This triggers the suspicion that the complex structure of the bipartite graph might be read off directly from the geometry of the Calabi--Yau.

In order to test this hypothesis, one should identify in the CY$_3$ geometry the dimer itself.
This is an open question, in the end related to the underlying reason for the coding of scaling dimensions in terms of angles.
In this letter, we take a first step toward realizing this goal.
Indeed, this small but crucial step will constitute the first investigation of dimer models and tilings from the differential geometry of the bulk Calabi--Yau.
In short, inspired by the proposal \cite{Feng:2005gw} of identifying the torus in which the dimer resides as part of the $\mT^3$ in the SYZ prescription of mirror symmetry \cite{Strominger:1996it}, we will attempt to find this torus explicitly.
Indeed, mirror symmetry~\cite{Strominger:1996it} can be thought of as $T$-duality once we exhibit the CY$_3$ as a supersymmetric torus fibration.
This suggests that we should regard the CY$_3$ as a special Lagrangian (SLag) torus fibration, identifying the $U(1)^2$ with the relevant $\mathbb{T}^2$ on which the dimer lives.
This suggests a way to metrically identify the torus, in particular allowing us to compute its complex structure parameter.

On the other hand, as discussed above, field theory arguments suggest this complex structure will take a certain value for each ${\cal N}=1$ SCFT, namely the one determined by the R-charges in a so-called isoradial embedding of the dimer.
It is thus natural to guess that the complex structure of the $\mathbb{T}^2$ identified in the geometry, which we denote as $\tau_G$, will match the complex structure computed in field theory, which we denote as $\tau_R$.
This comparison will be the heart of our present investigation.

Indeed, we find the match $\tau_G = \tau_R$ to be true for the spaces ${\mathbb C}^3$ and the conifold, as well as for symmetric orbifolds of these spaces.
Quite interestingly, this na\"{\i}vely expected equality of complex structures is not quite realized in general SCFTs as the geometrical torus is slightly different from the field theoretic torus.
This ``not quite'' is in fact remarkably fascinating.
We find that the Klein $j$-invariants of the two tori are tantalizingly close numerically.
Given the extremely complicated nature of the $j$-invariant, this numerical proximity is highly non-trivial.
The two $\tau$-parameters are bound \textit{not} to agree due to the fact that the field theory $\tau_R$, can be argued to be, generically, a transcendental number using the \textit{four exponentials conjecture} while the geometrical $\tau_G$, computed from the explicit metrics known in the literature, can be shown to be an algebraic number.
This leaves a very interesting open question as for the origin of this small mismatch.
In the remainder of the paper we will further explain the above ideas, in particular showing explicitly the SLags in the geometry, which is \textit{per se} an interesting mathematical problem, and demonstrate how they fail by a tiny amount to reproduce the field theory result.

This paper is organized as follows.
We begin with Section~\ref{sec:background}, in which we briefly review dimer models, their description in terms of branes and SLags.
In Section~\ref{sec:c3}, we then construct the SLag for ${\mathbb C}^3$ corresponding to the ${\cal N}=4$ super-Yang--Mills theory and find that $\tau_R = \tau_G$.
We next do the same for the Klebanov--Witten theory on the conifold in Section~\ref{sec:conifold}, and find agreement once again.
Section~\ref{sec:orbifolds} briefly comments on the concordance of $\tau_R$ and $\tau_G$ for orbifolds of ${\mathbb C}^3$ and the conifold.
Section~\ref{sec:Labc} examines the more general case of $L^{a,b,c}$ and $Y^{p,q}$ spaces.
We again verify that for the spaces $Y^{p,0}$ and $Y^{p,p}$, the agreement between $\tau_R$ and $\tau_G$ continues to hold.
This must be the case as these are the orbifolds that we discussed in the previous section.
However, for $L^{a,b,a}$ spaces there is a small mismatch between the two $\tau$ parameters.
Section~\ref{sec:conclusions} speculates on the origin of the mismatch and presents avenues for further investigation.

\section{Toric CFTs, five-branes, and SLags}
\label{sec:background}

As described in Section~\ref{sec:intro}, the theories we are interested in are encoded as \textit{dimer models}.
In a nutshell, dimer models are bipartite graphs drawn on a torus consisting on black and while nodes linked in a certain way.
Each face of the graph represents a gauge group $SU(N)$ of the corresponding gauge theory.
Links join black and white nodes and separate two faces;
these correspond to bifundamental fields charged under the gauge groups associated to the faces that they separate, with the definition of fundamental and antifundamental defined by the orientation of the genus one Riemann surface.
We number the links to label the fields.
The dimer models reflect the toric character of the theory expressed in a superpotential that can be written as $W = W_+ - W_-$, where each field appears exactly once in each of $W_+$ and $W_-$.
In the dimer, vertices correspond to the superpotential terms.
Going around the black nodes clockwise, the fields label a monomial whose trace appears in $W_+$.
Going around the white nodes counterclockwise, the fields label a monomial whose trace appears in $W_-$.
In this way, we capture the field content of the ${\cal N}=1$ SCFT and the terms in the superpotential.

\begin{figure}[h!]
\centering
\includegraphics[trim=10mm 0mm 0mm 0mm, clip, width=4.5in]{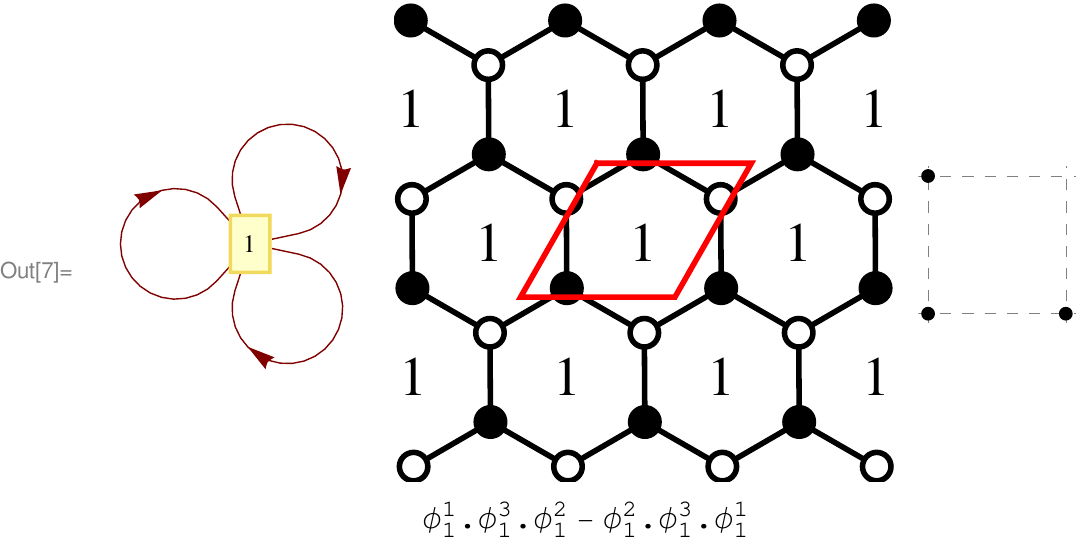}
\caption{{\sf The quiver, dimer and toric diagram for ${\cal N}=4$ super-Yang-Mills theory in 4-dimensions, corresponding to the toric Calabi--Yau threefold $\IC^3$.}}
\label{f:c3-dimer}
\end{figure}

The canonical example of $\cN=4$ super-yang-Mills theory is, in fact, representable as a dimer model.
In Figure \ref{f:c3-dimer}, we illustrate the above rules diagrammatically with this important case.
The ``clover quiver'' with the three adjoints $\phi^{1,2,3}_1$ is shown at the far left (the subscript 1 is to emphasize that all these three fields are charged under the group corresponding to face ``1'' in the dimer model, shown in the middle.
To the right, we include the planar toric diagram for $\IC^3$ for completeness.
Below the diagrams we show the famous trivalent superpotential.
To this theory and many more we shall shortly return.

\subsection{Physical origin of the dimer}
The existence of an underlying torus may at first seem mysterious and was initially forced upon us by the curious fact that for all toric quiver gauge theories we have
\begin{equation}
N_W - N_E + N_G = 0 ~.
\end{equation}
Here $N_W$ is the number of monomials in the superpotential, $N_E$ is the total number of fields, and $N_G$ is the total number of gauge group factors.
In the dimer, these are, respectively, the number of nodes, edges, and faces in the fundamental region.
Of course, we recognize this as the Euler relation for a genus one Riemann surface, whence the torus.
The physical motivation of the dimer construction, whereby explaining why AdS$_5$/CFT$_4$ should obey this topological condition was expounded in~\cite{Feng:2005gw}.
The answer turns out to be mirror symmetry.

We know that upon mirror symmetry, the original IIB setup of D$3$-branes at the tip of the Calabi--Yau cone ${\cal M}$ gets mapped to a system of intersecting D$6$-branes on the mirror Calabi--Yau ${\cal W}$.
The mirror ${\cal W}$ is given by the following complete intersection in $\IC[x,y,u,v,z]$:
\begin{equation}
u\,v=z ~, \qquad
P(x,\,y)=z ~, \qquad
z\in \mathbb{C} ~,
\end{equation}
where $P(x,\,y)$ is the Newton polynomial of the toric diagram of the $\cM$.
This polynomial is constructed starting from the toric diagram of the CY$_3$, which we recall is a collection of lattice two-vectors $\{ (p_i, q_i) \} \in \IC^2$.
Then the Newton polynomial is given by
\begin{equation}
P(x,y) = \sum_{\{{\rm points\ in\ toric\ diagram}\}}\,a_i\,x^{p_i}\,y^{q_i} ~,
\end{equation}
where the $a_i$ are complex numbers parameterizing the complex moduli of $\cW$, and hence the K\"ahler moduli of the original $\cM$.

\begin{figure}[h!]
(a)\includegraphics[trim=0mm 0mm 0mm 0mm, clip, width=4in]{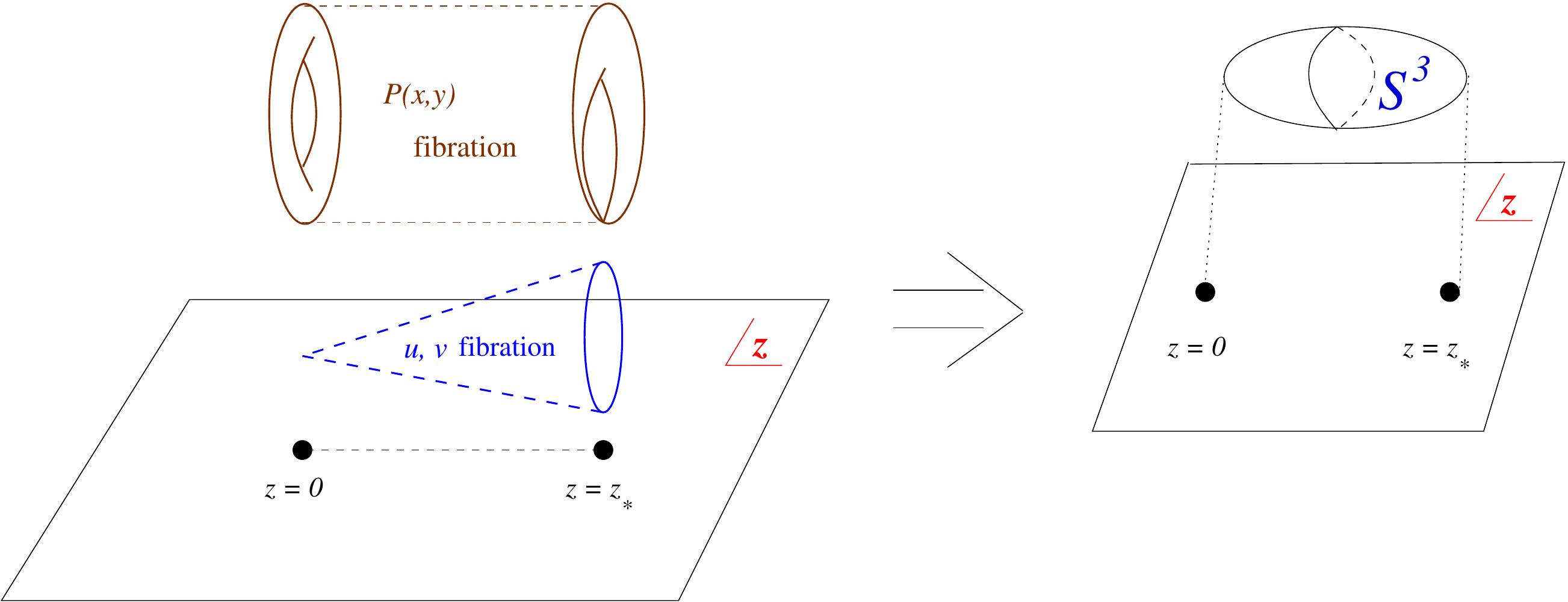}
(b)\includegraphics[trim=0mm 0mm 0mm 0mm, clip, width=2in]{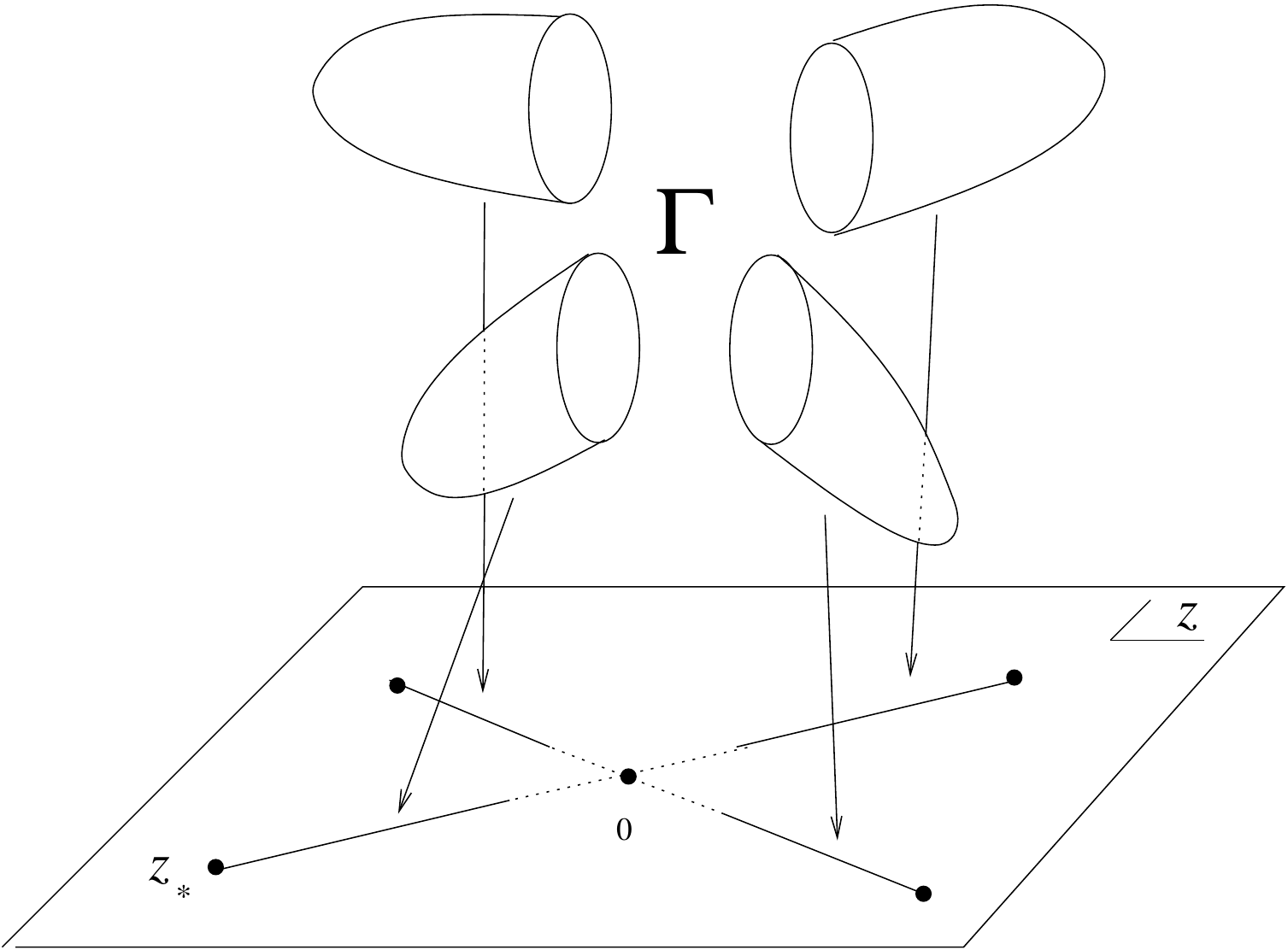}
\caption{{\sf (a) The mirror of the Calabi--Yau threefold as a double fibration over $\IC$.
(b) The $S^3$ cycles meet at the origin in the $z$-plane on a finite graph, which is the dimer model.
}}
\label{f:mirrorfibre}
\end{figure}

The mirror $\cW$ is in fact a double fibration over $\mathbb{C}$.
The equation $P(x,\,y)=z$ defines, for each point $z$, a certain Riemann surface $\Sigma_z$.
The other fibration contains an $S^1$ corresponding to $\{u,\,v\}\rightarrow \{e^{i\,\theta}\,u,\,e^{-i\,\theta}\,v\}$; obviously this $U(1)$ collapses at $z=0$.
The surface $\Sigma_z$ develops singularities at some critical points $z_* = z^{cr}_i$ where $\partial_x P= \partial_y P = 0$.
At these points $z_i^{cr}$, a one-cycle of $\Sigma_z$ pinches off.
Hence, over the segment on the $z$-plane joining $z=0$ and $z_i^{cr}$ there is a $U(1)^2$, which is pinching off at the ends in a certain way.
This is topologically an $S^3$ where the D$6$-branes are wrapped.
There will be one $S^3$ for each critical point of $\Sigma_z$, which meet at the point $z=0$.
This is illustrated in part (a) of Figure \ref{f:mirrorfibre}.
The dimer itself is then the intersection of these $S^3$ cycles at the origin of the $z$-plane, as some finite graph $\Gamma$; this is shown in part (b) of the figure.

From the expression of the Newton polynomial $P(x,y)$, it is clear that each monomial on $P$ specifies a $U(1)$ as $(x,\,y)\rightarrow (e^{-i\,q_i\,\theta}\,x,\,e^{i\,p_i\,\theta}\,y)$, which can be seen as follows.
The critical points $(x_cr,y_cr)$ satisfy
\begin{equation}
dP(x,y)|_{(x_{cr},y_{cr})} = 0 \quad \Longrightarrow \quad
\sum_i a_i\,p_i\,y_{cr}\,x_{cr}^{p_i}\,y_{cr}^{q_i}=0 ~, \qquad
\sum_i \, a_i\,q_i\,x_{cr}\,x_{cr}^{p_i}\,y_{cr}^{q_i}=0 ~.
\end{equation}
Thus, for each monomial we can find a critical point by setting $-q_i\,x_{cr}+p_i\,y_{cr}=0$, hence finding on the $(x,\,y)$ plane a complex line with a slope given by $(p_i,\,q_i)$.
Since $x$ and $y$ are complex variables, there is a $\mathbb{T}^2$ associated to their phases in the natural way $(x,\,y)=(r_x\,e^{i\,\theta_x},\,r_y\,e^{i\,\theta_y})\rightarrow (\theta_x,\,\theta_y)$.
Each monomial therefore defines a one-cycle winding on $\mathbb{T}^2$ as specified by $(p_i,\,q_i)$, and this one-cycle serves as base for a cylinder which develops around each critical point of $P$.

As discussed above, the critical points in turn correspond to D$6$-branes.
This thus serves as a natural way to identify their winding on $\Sigma_0$.
In fact, recalling that the so-called $(p,\,q)$-web is the graph dual to the toric diagram, as
it was realized as far back as \cite{Aharony:1997bh}, $\Sigma_0$ is nothing but the thickened $(p,\,q)$-web associated to the CY$_3$.
In the language of \cite{Feng:2005gw}, the $(p,q)$-web is the spine of the am{\oe}ba projection of $\Sigma_0$.
Once we have identified the D$6$-branes on $\Sigma_0$, it is natural to consider their projection to the $\mathbb{T}^2$ defined by $(\theta_x,\,\theta_y)$.
This defines the so-called alga map, and it indeed shows the dimer in an explicit way \cite{Feng:2005gw}.

It was later understood~\cite{Imamura:2007dc} that the construction in~\cite{Feng:2005gw} can be related to a brane tiling of a $\mathbb{T}^2$.
This can be heuristically understood starting with the original IIB configuration of $N$ D$3$-branes probing a toric CY$_3$ and performing two $T$-dualities along two coordinates of the toric fiber.
As the toric fibers shrink somewhere on the base, these $T$-dualities produce a certain arrangement of NS$5$-branes winding around the torus.
In turn, the D$3$-branes will map into D$5$-branes wrapping the torus.
However, as is well-known, NS$5$-branes and D$5$-branes must join into $(p,\,q)$ five-branes running at certain angles in the web plane so as to preserve supersymmetry.
Thus, the system of NS$5$-branes and D$5$-branes become a tiling of the $\mathbb{T}^2$.
It turns out that the $(p,\,q)$ five-branes separating diverse regions of the tiling do form the $(p,\,q)$-web of the CY$_3$.

Indeed, these two pictures above are related by $T$-duality:
by $T$-dualizing the collapsing $S^1$ in the $\{u,\,v\}$ plane in the construction of~\cite{Feng:2005gw}, the D$6$-branes get mapped to D$5$-branes.
But, as the $S^1$ is collapsing, $T$-duality produces a configuration of NS$5$-branes following the $(p,\,q)$-web, thus recovering the picture in~\cite{Imamura:2007dc}.
For excellent reviews with further details we refer the reader to~\cite{Kennaway, Yamazaki:2008bt}.
As described in~\cite{Yamazaki:2008bt}, the gauge theory can be read off from the brane system directly.

\subsection{R-charges and $\tau_R$}
Our discussions above identify the dimer topologically as living on a $\mT^2$ part of the $\mT^3$ fibration in mirror symmetry.
We can further fix this torus, and this distinguishes the so-called {\it isoradial embedding} of the dimer.

Now, the R-charges of the fields in the theory are determined through the standard $a$-maximization procedure~\cite{iw}.
First of all, we demand that the sum of the R-charges of fields that appear in a monomial in the superpotential is two; this is just so that the Lagrangian is well-defined when written as a superspace integral:
\begin{equation}\label{sumR=2}
\sum_{\mbox{around each vertex}} R_i = 2 ~.
\end{equation}
Secondly, associated to each face of the dimer is a gauge group, whose $\beta$ function is
\be\label{betaG=0}
\beta_G = \frac{3N}{2(1-\frac{g^2 N}{8\pi^2})} \left( 2 - \sum_{\mbox{around each face}} (1 - R_i) \right) ~,
\ee
and which vanishes for conformality.
The $R_i$ in the expression is an R-charge of a field in fundamental or antifundamental representations of $G$.\footnote{
As usual, adjoints are regarded as both a fundamental and an antifundamental field and counted twice.}
Subject to these constraints, we maximize the central charge
\be
a = \frac{3}{32} \left( 3 \sum_i (R_i - 1)^3 - \sum_i (R_i - 1) \right) ~,
\ee
whereby fixing the values of all the R-charges of all the fields in the gauge theory.

With these R-charges, we can fix the dimer (and hence the torus).
First, we draw lines drawn from the center of each face to a vertex to be of unit length.
This is called an \textit{isoradial embedding} because now each face is inscribed by a unit circle.
There is a moduli space of such embeddings.

Now, the R-charge of a field, which is an edge in the dimer, is geometrically interpreted as the angle at the origin of each unit circle subtended by the equilateral triangle defined by the edge: $\theta_i = \pi R_i$.
In this way, condition \eqref{sumR=2} is just that as we circumnavigate any of the vertices the angle sum is $2\pi$, and thus the dimer is truly planar.
Similarly, condition \eqref{betaG=0} dictates that lines joining all centers of faces prescribe rhombi and we have a rhombus tiling of the plane.
Furthermore, the length of an edge is easily seen as $2\cos\frac{\pi}{2} R_i$.

With $R_i$ being the R-charge of the associated field as determined by $a$-maximization, we select a particular dimer from the moduli space of the isoradial dimers~\cite{hv}.
The complex structure parameter of the dimer drawn in this way is $\tau_R$.
We will hence forth use this particular complex parameter from the field theory.\footnote{
A bipartite graph on a torus $\mathbb{T}^2$ can be encoded in terms of a Belyi pair, consisting of an elliptic curve $\Sigma_1$ together with a holomorphic map $\beta: \Sigma_1 \to \mathbb{P}^1$ branched over three points on the $\mathbb{P}^1$~\cite{Jejjala:2010vb}.
The elliptic curve that is the source for this map has its own complex structure parameter $\tau_B$.
For $\mathbb{C}^3$ and the conifold and orbifolds thereof, $\tau_R = \tau_B$~\cite{Jejjala:2010vb}.
This is not, however, true generally~\cite{Hanany:2011ra}.}

\subsection{Special Lagrangian fibrations and $\tau_G$}
The object of our investigation is to explicitly identify the $\mathbb{T}^2$ on which the dimer lives.
As briefly reviewed above, this $\mathbb{T}^2$ is topologically identified with the complex structure of the Newton polynomial in the mirror, or, equivalently, with the $\mathbb{T}^2$ where the brane tiling lives.
In order to make the identification more precise, let us focus on the mirror symmetry transformation of~\cite{Feng:2005gw}.

As described in~\cite{Strominger:1996it}, mirror symmetry on toric Calabi--Yau threefolds can be understood as fiberwise $T$-duality on the original CY$_3$.\footnote{
To the best of our knowledge, this is however still a heuristic picture in that its precise mathematical characterization remains to be fully understood. Besides, our Calabi--Yau is non-compact, which adds a further subtlety.}
Following this inspiration, in our non-compact setup, we are instructed to regard the CY$_3$ as a special Lagrangian (SLag) fibration.
Recall that the SLag cycle $L$ in a CY$_3$ is a middle-dimensional manifold (\textit{i.e.}, a three-cycle)
satisfying
\begin{equation}
{\rm Im}\,({\rm P}_L[\Omega]\,)=0 ~, \qquad
{\rm P}_L[\omega]=0 ~,
\end{equation}
where $\Omega$ and $\omega$ are, respectively, the holomorphic three-form and the K\"ahler form of the CY$_3$ and ${\rm P}_L$ denotes the pull-back onto the SLag cycle.
It is then natural to regard the CY$_3$ itself as a fibration
\begin{equation}
f:\,{\mathrm CY}_3\,\rightarrow \mathbb{R}^3
\end{equation}
over $\mathbb{R}^3$ such that each fiber is a SLag cycle.

Our geometries, being toric, admit the action of an $U(1)^3$.
However, only a $U(1)^2$ subgroup will leave invariant $\Omega$ and $\omega$.
It is then natural to concentrate on SLag fibrations invariant under this $U(1)^2$.
Then, we can piece together the pictures of~\cite{Feng:2005gw} and~\cite{Imamura:2007dc}:
$T$-dualizing this $U(1)^2$ will lead to the brane tiling of~\cite{Imamura:2007dc}, while a further $T$-duality will be analogous to mirror symmetry and take us to the picture in~\cite{Feng:2005gw}.
This is in fact summarized in Figure~82 of~\cite{Yamazaki:2008bt}.
This further suggests that the $U(1)^2$ defining the SLag also defines the $\mathbb{T}^2$ on which the dimer lives, the object of our primary interest.

Due to the $U(1)^2$ invariance, we can be more precise in defining our SLag fibration, which is defined in terms of the moment maps $\mu_i$ of the two $U(1)$ actions as $f=(f_0,\,\mu_1,\,\mu_2)\in\mathbb{R}^3$.
The $f_0$ is a ``generalized moment map'' for $\Omega$, appropriately chosen so that it ensures that the fibration is SLag; we will define this more precisely for explicit examples later.

The $U(1)^2$ defining the SLag will generically collapse in a certain way on $\mathbb{R}^3$.
The generic fiber over a collapsing locus is topologically $\mathbb{R}^+\times \mathbb{T}^2$.
Seen on $\mathbb{R}^3$, these form a set of lines supported on a $\mathbb{R}^2$ (on the plane $x_1=0$ in $\mathbb{R}^3$) that in fact coincides with the $(p,\,q)$-web on the Calabi--Yau.
The lines forming the web meet at a single point at the origin of $\mathbb{R}^2$, where the fiber is in fact metrically a torus over $\mathbb{T}^2$.
At this point, the collapsing torus coincides with the $U(1)^2$, which leaves the SLag invariant.
Hence, it is the point we will be most interested in understanding.
This $\mathbb{T}^2$ is naturally identified with the elliptic curve that supports the dimer.
By pulling back the CY$_3$ metric to this $\mathbb{T}^2$, we will be able to compute its associated complex structure, which we will call $\tau_G$, and which we will use to compare with the many field theory $\tau_R$ computed in~\cite{Hanany:2011bs}.

In summary, given the toric CY$_3$, we follow the following algorithm:
\begin{itemize}
\item find the metric where possible;
\item explicitly identify the SLag from the coordinates;
\item find the $U(1)^2$-invariant part of the SLag from the moment maps; this should be the torus on which the dimer lives;
\item pull back the metric to this torus and compute its complex parameter $\tau_G$
\item compare with $\tau_R$ from the isoradial dimer.
\end{itemize}

\section{Example: $\mathbb{C}^3$}
\label{sec:c3}
Let us begin with the simplest example, $\IC^3$.
The dimer model is that of $\cN=4$ super-Yang-Mills, as described in Figure \ref{f:c3-dimer}.
In a physics related context, a detailed discussion can be found in~\cite{Morrison:2010vf}.

We now follow the prescription of \cite{HL}.
Letting the complex coordinates of $\IC^3$ be $z_{1,2,3}$, the holomorphic three-form and the K\"ahler form are simply
\begin{equation}
\Omega = dz_1\wedge dz_2\wedge dz_3 ~, \qquad
\omega=\frac{i}{2}\,\sum_{i=1}^3\,dz_i\wedge d\bar{z}_i ~.
\end{equation}
The $U(1)^3$ toric fiber is generated by the rotations $z_i\rightarrow e^{i\,\theta_i}\,z_i$.
However, it is clear that only a two-dimensional subspace, namely a $U(1)^2$, will leave invariant both $\Omega$ and $\omega$.
We can choose
\begin{eqnarray}
\nn
U(1)_1\,:\,&& (z_1,\,z_2,\,z_3)\,\rightarrow\, (e^{i\,\theta_1}\,z_1,\,z_2,\,e^{-i\,\theta_1}\,z_3)\\
U(1)_2\,:\,&& (z_1,\,z_2,\,z_3)\,\rightarrow\, (z_1,\,e^{i\,\theta_2}\,z_2,\,e^{-i\,\theta_2}\,z_3)
\end{eqnarray}
as the $U(1)^2$ action.
Note that the fixed points of the $U(1)^2$ are $z_1=z_3=0$, where $U(1)_1$ collapses; $z_2=z_3=0$, where $U(1)_2$ collapses, and $z_1=z_2=0$, where $U(1)_1-U(1)_2$ collapses.

The moment maps associated to $U(1)_{1,2}$ are, respectively,
\begin{equation}
\mu_1=|z_1|^2-|z_3|^2 ~, \qquad
\mu_2=|z_2|^2-|z_3|^2 ~.
\end{equation}
Then, the SLag fibration reads
\begin{equation}
f=({\rm Im}\,(z_1\,z_2\,z_3),\,|z_1|^2-|z_3|^2,\,|z_2|^2-|z_3|^2) ~,
\end{equation}
where the first entry has been chosen so that every fiber is SLag.
Indeed, $f$ maps the non-compact CY$_3$, here just $\IC^3$, to $\IR^3$ and the fibers are $\IR \times U(1)^2$.

Let us now look at the fixed loci, as mentioned above.
These are the following:
\begin{itemize}
\item $\mathcal{S}_1$: $U(1)_1$ collapses and
the fixed loci are $z_1=z_3=0$.
Thus,
\begin{equation}
f=(0,\,0,\,x) ~, \qquad
x\in \mathbb{R}^+ ~;
\end{equation}
\item $\mathcal{S}_2$: $U(1)_2$ collapses and
the fixed loci are $z_2=z_3=0$.
Thus,
\begin{equation}
f=(0,\,x,\,0) ~, \qquad
x\in \mathbb{R}^+ ~;
\end{equation}
\item $\mathcal{S}_3$: $U(1)_1-U(1)_2$ collapse and
the fixed loci are $z_1=z_2=0$.
Thus,
\begin{equation}
f=(0,\,-x,\,-x) ~, \qquad
x\in \mathbb{R}^+ ~.
\end{equation}
\end{itemize}

Defining the set $PQ=\mathcal{S}_1\,\cup\, \mathcal{S}_2\,\cup\, \mathcal{S}_3$, we can readily construct the $(p,\,q)$-web for $\mathbb{C}^3$.
This is just given by the three directional vectors in the above, \textit{viz.}, $(0,0,1)$, $(0,1,0)$ and $(0,-1,-1)$.
Indeed, this is just the planar graph dual to the toric diagram given in Figure \ref{f:c3-dimer}, illustrating our discussion above that the collapsing cycles should give the spine of the am{\oe}ba.

Topologically, every SLag fiber is a cone over a $\mathbb{T}^2$.
However, there is a special fiber at the center $z_1=z_2=z_3=0$ of the $(p,\,q)$-web which is metrically a cone over a $\mathbb{T}^2$.
This follows because at the origin of the web $f=0$, a scaling symmetry  $f\rightarrow \lambda \,f$ appears.
As this is the point over which all legs of the web meet, this is the fiber whose $\mathbb{T}^2$ is the subject of our investigation.
As explained above, this is where the dimer should reside.
Let us denote this special fiber as $L_0$.
It is fairly easy to see explicitly.

Writing $z_i=r_i\,e^{i\psi_i}$, the origin of the web is at
\begin{equation}\label{centerC3}
\psi_1+\psi_2+\psi_3=0,\,\pi ~; \qquad
r_1=r_2=r_3=r ~.
\end{equation}
In order to manifestly exhibit the conelike structure, we now note that the $\mathbb{C}^3$ metric can be written as
\be
ds^2=\sum\limits_{i=1}^3 ( dr_i^2+r_i^2\,d\psi_i^2 ) ~.
\ee
Introducing $\rho=\sqrt{3}\,r$ and using \eqref{centerC3}, the pull-back of the metric to $L_0$ is
\begin{equation}
ds^2=d\rho^2+\frac{1}{3}\,\rho^2\,\Big[ d\psi_1^2+d\psi_2^2+(d\psi_1+d\psi_2)^2\Big] ~,
\end{equation}
where the cone structure of $\IC^3$ is now apparent: $\rho$ is now the lateral side of the cone, and the base is the $\mathbb{T}^2$ whose metric is
\begin{equation}\label{T2-C3}
ds_{\mathbb{T}^2}^2=\frac{1}{3}\,\Big[ d\psi_1^2+d\psi_2^2+(d\psi_1+d\psi_2)^2\Big] ~.
\end{equation}

Our proposal, as discussed in the previous section, is that this $\mT^2$ is where the dimer lives.
What is its complex structure?
We can easily determine it and refer the reader to Appendix~\ref{sec:sexytau} on how to find the $\tau$-parameter in general.
Our metric in \eqref{T2-C3} corresponds to the case in the appendix where $A=B=C=\frac{2}{3}$ and we readily obtain the complex structure
\begin{equation}
\tau_G=\frac{1}{2}\,\Big(1+i\,\sqrt{3}\Big)=e^{i\frac{\pi}{3}} ~.
\end{equation}
Very nicely, this exactly matches $\tau_R$ from the isoradial embedding (\textit{cf.}~\cite{Jejjala:2010vb}).

\subsection{An alternative point of view}
In order to better understand the physics as well as the mathematics, let us do the same computation for $\mathbb{C}^3$ in a slightly different language, which we will later use to compute more complicated examples.
The space $\mathbb{C}^3$ can be thought of as a cone over $S^5$, which in turn is a $U(1)$ bundle over $\mathbb{P}^2$.
In order to see this, let us introduce
\begin{equation}\label{ang-C3}
z_0=r\,\cos\phi_1\,e^{i\chi} ~, \qquad
z_1=r\,\sin\phi_1\,\cos\frac{\phi_2}{2}\,e^{i\frac{2\,\chi+\psi+\phi_3}{2}} ~, \qquad
z_2=r\,\sin\phi_1\,\sin\frac{\phi_2}{2}\,e^{i\frac{2\,\chi+\psi-\phi_3}{2}} ~,
\end{equation}
so that $\sum\limits_{i=0}^2 |z_i|^2=r^2$.
The homogeneous coordinates on $\mathbb{P}^2$ are given by
$\hat{z}_1=\frac{z_1}{z_0}$ and $\hat{z}_2=\frac{z_2}{z_0}$.
The range of the coordinates here is $\phi_1\,\in\,[0,\,\frac{\pi}{2}]$, $\phi_2\,\in\,[0,\,\pi]$, and $\phi_3\,\in\,[0,\,2\,\pi]$, while $\psi\,\in\,[0,\,4\,\pi]$ and $\chi\,\in\,[0,\,2\,\pi]$.

By substituting the angular expressions \eqref{ang-C3} into the standard expressions for the $\mathbb{C}^3$ metric, K\"ahler form, and holomorphic form it is straightforward to construct the corresponding expressions in angular coordinates.
Specifically, we have that
\begin{equation}
ds^2=(d\chi-A)^2+ds_{\mathbb{P}^2}^2 ~,
\end{equation}
where
\begin{equation}
A=-\frac{1}{2}\,\sin^2\phi_1\,\Big(d\psi+\cos\phi_2\,d\phi_3\Big)
\end{equation}
and
\begin{equation}\label{metricP2}
ds_{\mathbb{P}^2}^2=d\phi_1^2+\frac{1}{4}\,\sin^2\phi_1\,\Big[\,\cos^2\phi_1\,\Big(d\psi+\cos\phi_2\,d\phi_3\Big)^2+d\phi_2^2+\sin^2\phi_2\,d\phi_3^2\,\Big] ~.
\end{equation}

Na\"{\i}vely, the relevant $U(1)^2$ is given in terms of $\{\partial_{\phi_3},\,\partial_{\psi}\}$.
In fact, the center of the SLag is now given by $\phi_2=\frac{\pi}{2}$, $\phi_1=\frac{1}{2}\,\arccos\frac{1}{3}$ and $\psi=-3\,\chi$.
Taking the pull-back of the metric we find
\begin{equation}
ds^2_c=\frac{1}{6}\,\Big(d\phi_3^2+3\,d\chi^2\Big) ~,
\end{equation}
which clearly does not yield the correct $\tau$.
This is due to somewhat subtle global issues that we have not taken into account.

Let us first start by noticing that for fixed $\phi_1$, in the $\mathbb{P}^2$ base we find, using \eqref{metricP2}, locally an $S^3$.
The $\mathbb{T}^2$ is written in terms of the $\varphi_i$ angles defined by
\begin{equation}
\psi=2\,\varphi_1+\varphi_2 ~, \qquad
\phi_3=\varphi_2 ~,
\end{equation}
so that
\begin{equation}
\partial_{\varphi_1}=2\,\partial_{\psi} ~, \qquad
\partial_{\varphi_2}=\partial_{\psi}+\partial_{\phi_3} ~.
\end{equation}
See, \textit{e.g.}, Section 4 of \cite{Martelli:2004wu}.

Writing the $z_i$ as $z_i=r_i\,e^{i\,\psi_i}$, the toric $\mathbb{T}^3$ is nothing but the three $\psi_i$ coordinates.
In terms of the $\varphi_i$, the transformation reads
\begin{equation}
\psi_1=\chi ~, \qquad
\psi_2=\chi+\varphi_1+\varphi_2 ~, \qquad
\psi_3=\chi+\varphi_1 ~,
\end{equation}
or, in matrix form
\begin{equation}
\left(\begin{array}{c} \psi_1 \\ \psi_2 \\ \psi_3 \end{array}\right) = M\,\left(\begin{array}{c} \chi \\ \varphi_1 \\ \varphi_2 \end{array}\right) ~, \qquad
M = \left( \begin{array}{ccc} 1 & 0 & 0 \\ 1 & 1 & 1 \\ 1 & 1 & 0\end{array}\right) ~.
\end{equation}
Note that ${\rm det}M=-1$, so this is an $SL(3,\,\mathbb{Z})$ transformation.\footnote{
The overall sign is just due to orientation.
Upon simply sending $\chi$ to $-\chi$ we would recover the unit determinant.}
Thus, we see that the set of coordinates $(\chi,\,\varphi_1,\,\varphi_2)$ are a good global basis for the toric $\mathbb{T}^3$ as well as the $\mathbb{T}^2$.

In terms of these coordinates the metric at the center of the SLag is
\begin{equation}
ds_c^2=\frac{2}{9}\,\Big(d\varphi_1^2+d\varphi_2^2+d\varphi_1\,d\varphi_2\Big) ~,
\end{equation}
which leads to the expected $\tau_G = \tau_R = \exp( \frac{\pi i }{3})$.

\section{Example: the conifold}
\label{sec:conifold}
\begin{figure}[h!t!!]
\centering
\includegraphics[trim=10mm 0mm 0mm 5mm, clip, width=4.5in]{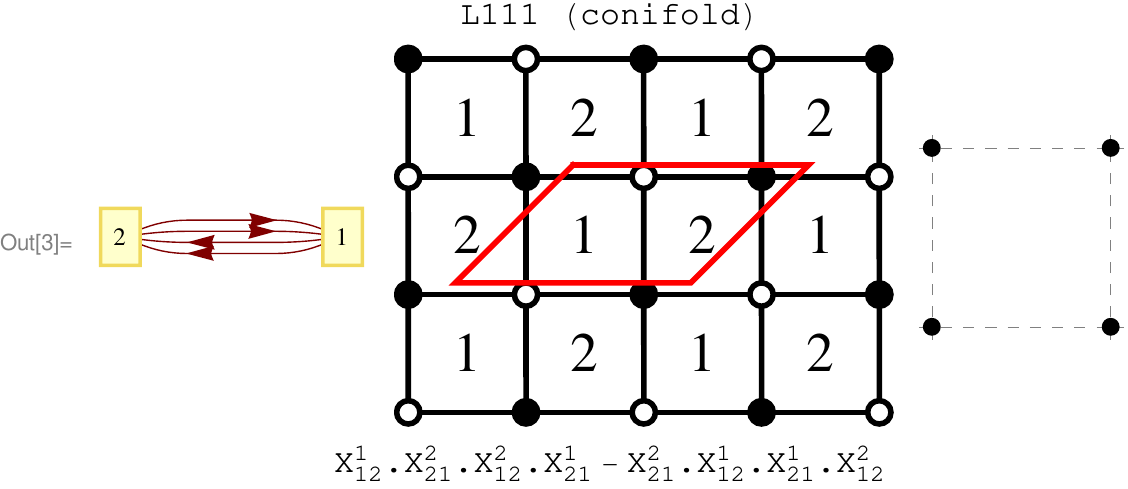}
\caption{{\sf The quiver, dimer and toric diagram for the conifold theory.
The quartic superpotential is given underneath.}}
\label{f:coni}
\end{figure}

Encouraged by the success of the matching for the simplest case of $\IC^3$,
let us now look at a more involved example, namely the conifold theory.
For reference, the quiver, dimer, and toric diagram are given in Figure \ref{f:coni}.

The metric for the conifold is the earliest known example for a (non-compact) Calabi--Yau manifold \cite{Candelas:1989js}:
\begin{equation}
ds^2=dr^2+\frac{r^2}{9}\,g_5^2+\frac{r^2}{6}\,\Big(\sum_{i=1,\,2}\,e_{\theta_i}^2+e_{\phi_i}^2\Big) ~,
\end{equation}
where we have defined
\begin{equation}
g_5=d\psi_R-\cos\theta_1\,d\phi_1-\cos\theta_2\,d\phi_2 ~, \qquad
e_{\theta_i}=d\theta_i ~, \qquad
e_{\phi_i}=\sin\theta_i\,d\phi_i ~.
\end{equation}
Here $\phi_i\,\in\,[0,\,2\,\pi]$, $\theta_i\,\in\,[0,\,\pi]$, and $\psi_R\,\in\,[0,\,4\,\pi]$.

By forming the following combinations
\begin{equation}\label{coni-z}
e_1=e^{i\psi_R}\,\Big(dr+i\,\frac{r}{3}\,g_5\Big) ~, \qquad
e_2=\frac{r}{\sqrt{6}}\,\Big(e_{\theta_1}+i\,e_{\phi_1}\Big) ~, \qquad
e_3=\frac{r}{\sqrt{6}}\,\Big(e_{\theta_2}+i\,e_{\phi_2}\Big) ~,
\end{equation}
we see that the metric is just $ds^2=\sum\limits_{i=1}^3\,|e_i|^2$, so that the K\"ahler and top holomorphic forms take the standard $\mathbb{C}^3$ form.
In particular, the K\"ahler form is
\begin{equation}
\omega=\frac{i}{2}\,\sum\,e_i\wedge \bar{e}_i=\frac{r}{6}\,\Big( 2\,dr\wedge g_5+r\,e_{\theta_1}\wedge e_{\phi_1}+e_{\theta_2}\wedge e_{\phi_2}\Big) ~.
\end{equation}

Therefore, as we have done above, we can find the $U(1)^2$ invariant part of the $\mT^3$ with the action of the $U(1)^2$ generated by $\{\partial_{\phi_1},\,\partial_{\phi_2}\}$.
It is not difficult to construct their moment maps, which read
\begin{equation}
\mu_{\phi_1}=-\frac{r^2}{6}\,\cos\theta_1 ~, \qquad
\mu_{\phi_2} =- \frac{r^2}{6}\,\cos\theta_1 ~.
\end{equation}

The SLag sits at $\mu_{\phi_i}=x_i$, where $x_i$ are real constants.
However, since we are interested on the center of the $(p,\,q)$-web, we can set $x_i=0$, that is, $\theta_i=\frac{\pi}{2}$.
Furthermore, we have $\Omega\sim e^{i\,\psi_R}$ when substituting \eqref{coni-z} into $e_1\wedge e_2\wedge e_3$.
This just shows that $\psi_R$ is identified with the field theory R-symmetry.
The SLag condition demands that $\psi_R=0$.

In summary then, we find, upon taking $\psi_R=0$, $dr=0$, and $\theta_1=\theta_2=\frac{\pi}{2}$ both moment maps vanish while ${\rm Im}\, \Omega={\rm P}_L[\omega]=0$.
Thus, this corresponds to the $U(1)^2$-invariant SLag at the center of the web where the dimer should live.
The pull-back of the metric there is
\begin{equation}
\label{conifoldSLag}
ds_{\mT^2}^2=\frac{r^2}{6}\,\Big(d\phi_1^2+d\phi_2^2\Big) ~,
\end{equation}
which is just a square torus, so we find $\tau_G=i$.
Note that in this case there are no further global issues, as $\{\partial_{\phi_i}\}$ do indeed cover a torus.\footnote{
Strictly speaking, the globally well-defined Killing vectors are $\partial_{\phi_i}+\partial_{\psi_R}$, but for our purposes we can just consider $\partial_{\phi_i}$ since $\psi_R=0$.}
Once again, the geometric complex structure $\tau_G$ so obtained is the same as the complex structure $\tau_R$ of the isoradial dimer with edge lengths fixed by the R-charges~\cite{Jejjala:2010vb}.

\section{Orbifolds}
\label{sec:orbifolds}

An immediate consequence for the above examples is that we can directly compute the $\tau_G$ for all their orbifolds.
As the local form for the metric will be unaffected by the orbifolding procedure, we can just borrow the study of the SLag submanifolds from the unorbifolded geometries.
However, the global analysis will be different, as orbifolding will change the periodicity of the angles.
The simplest example is the non-chiral orbifold of the conifold, which corresponds to taking $\phi_1\,\in\,[0,\,\pi]$.
We can borrow the above result (\ref{conifoldSLag}) provided we redefine $\phi_2=\frac{\tilde{\phi}_2}{2}$ so that $\tilde{\phi}_2\,\in\,[0,\,2\,\pi]$ so that we can use the formul\ae\ in the appendix.
The metric is then

\begin{equation}
ds_{\mT^2}^2=\frac{r^2}{6}\,\Big(d\phi_1^2+\frac{1}{4}\,d\tilde{\phi}_2^2\Big) ~,
\end{equation}
which leads to $\tau_G=2\,i$.
This is precisely the expected result along the lines of \cite{Jejjala:2010vb,Hanany:2011ra}.
Moreover, from here, it is immediate that the orbifold pattern for $\tau_G$ will follow the dimer pattern for $\tau_R$, and so if the unorbifolded space agrees, all higher orbifolds will also show the $\tau_R=\tau_G$ agreement.

\section{$Y^{p,q}$ and $L^{a,b,c}$ manifolds}
\label{sec:Labc}

The examples above indicate that global issues should be relevant when tackling more involved manifolds, such as the $Y^{p,q}$ spaces constructed in~\cite{Martelli:2004wu} and the $L^{a,b,c}$ spaces constructed in~\cite{Cvetic:2005ft, Martelli:2004wu}.
Instead of considering local expressions for the metrics where in the above we encountered subtleties involving the well-definedness and periodicity of the angular coordinates, we will use the more powerful approach based on the symplectic structure of these manifolds that was developed in~\cite{Martelli:2005tp}.

\subsection{Symplectic coordinates}
The key observation is that, upon introducing suitable coordinates $\{y^i,\,\phi_i\}$, the metric can be encoded in terms of a symplectic potential $G=G(y)$ such that
\begin{equation}
ds^2=G_{ij}\,dy^i\,dy^j+G^{ij}\,d\phi_i\,d\phi_j ~,
\end{equation}
where
\begin{equation}
G_{ij}=\frac{\partial^2\,G}{\partial y^i\,\partial y^j} ~.
\end{equation}
Furthermore, the K\"ahler form reads
\begin{equation}
\omega = dy^i\wedge d\phi_i ~,
\end{equation}
while the top holomorphic form is
\begin{equation}
\Omega = e^{i\,\phi_1}\cdots ~.
\end{equation}

We must set $\phi_1=0$ so that ${\rm Im}\,\Omega=0$, while the $U(1)^2$ will be generated by $\partial_{\phi_2}$ and $\partial_{\phi_3}$.
These are globally well defined, and the associated moment maps are $\mu_2=y_2$ and $\mu_3=y_3$, respectively.
The center of the web will sit at $y_2=y_3=0$.
Thus, in all, the $U(1)^2$-invariant sub-torus of the SLag at the origin will be given by the metric:
\begin{equation}
ds_{\mathbb{T}^2}=G^{IJ}(y_2=y_3=0)\,d\phi_I\,d\phi_J ~, \qquad
I,\,J=2,\,3 ~.
\end{equation}
We shall then need to compute the complex structure $\tau_G$ of this torus.

We are left with the rather non-trivial task of constructing the symplectic potentials $G$.
Luckily this problem has been solved in~\cite{Martelli:2005tp} and~\cite{Oota:2005mr}, where it is shown that $G$ can be entirely constructed from the toric data and must be of the form
\be\label{Gpot}
G=G^{{\rm can}}+G^{{\rm b}}+g ~,
\ee
where $G^{{\rm can}}$ is related to the canonical part of the metric as in~\cite{guillemin}, $G^{{\rm b}}$ is associated to Reeb vector moduli described in~\cite{Martelli:2005tp}, and $g$ is a remainder function that is a homogeneous and degree one rational function in the $y$ coordinates.
In particular, $g$ satisfies a Heun equation found in~\cite{Oota:2005mr}.
Note that for the geometric analog of $a$-maximization, \textit{viz.}\ volume minimization, the $g$ drops out and plays no r\^ole.
As we shall see, however, in the case at hand, it has crucial significance.

\subsection{Warmup: $Y^{p,q}$}
Let us begin with the simpler case of the $Y^{p,q}$ spaces.
The dimer model and hence the quiver and superpotential are nicely given in \cite{Franco:2005rj} and the isoradial $\tau_R$ were listed for a few cases in \cite{Hanany:2011bs}.
As one can see, the theories are already very complicated, and $\tau_R$ typically lives in highly non-trivial field extensions of $\mQ$.

Let us now use the notation of \cite{Oota:2005mr} and use the toric diagram defined by the following outward normal lattice vectors --- one can check that these vectors are actually co-planar, as required for a Calabi--Yau:\footnote{We emphasize that we have given the normals rather than the actually vectors of the toric diagram, \textit{i.e.}, we are describing the $(p,\,q)$-webs.}
\begin{equation}\label{Ypq}
\{
v_1 = (1,-1,-p) ~,
v_2 = (1,0,0) ~,
v_3 = (1,-1,0) ~,
v_4 = (1,-2,-p+q)
\} ~.
\end{equation}
Moreover, one defines the quantities
\begin{equation}
\ell = \frac{q}{3 q^2 - 2 p^2 + p \sqrt{4 p^2 - 3 q^2}} ~,
B = \{3,-3,-\frac32 (p-q+\frac{\ell}{3})\} ~,
v_5 = B - v_1 - v_3 ~,
v_6 = - v_2 - v_4 ~,
\end{equation}
where $B$ is the Reeb vector after $Z$-minimization.\footnote{As a \textit{caveat lector}, we remark that in \cite{Martelli:2005tp} the coordinates
$\{(1,0,0), (1-,p-q-1,p-q),(1,p,p), (1,1,0)\}$ are used; we can see that the two are related by an $SL(3;\mZ)$ transformation.
The ones in \eqref{Ypq} can be taken to these by, for example, $A = {\scriptsize \left( \begin{array}{ccc}  1 & 0 & 0 \\
 -1 & -1 & 1 \\
 0 & q-p & p-q-1 \end{array} \right)}$.}

Using these data, we can readily determine the symplectic potential to be
\begin{equation}
G(y_1,y_2,y_3) =
\frac12 \sum\limits_{i=1}^6 A \cdot v_i \cdot \left( y_1, y_2, y_3 \right)^T
\log \left| A \cdot v_i \cdot \left( y_1, y_2, y_3 \right)^T \right| ~,
\end{equation}
where we have used the matrix $A$ to rotate to \cite{Martelli:2005tp}.
We stress that this is the form for the full symplectic potential as found in  \cite{Oota:2005mr}, including in particular the remainder function $g$.
The latter is associated to the $v_5$ and $v_6$ vectors, and so dropping them would only lead to the $G^{{\rm can}}+G^{{\rm b}}$ piece.

From this, as explained above, we can find the metric of the required torus to be
the $I=J=2,3$ submatrix of
\begin{equation}
G^{IJ} =
\left. \left( \frac{\partial G}{\partial y_i \partial y_j} \right)^{-1}
\right|_{y_2 = y_3 = 0} ~.
\end{equation}
From this expression and Appendix \ref{sec:sexytau}, we can determine the $\tau_G$.

Let us tabulate some of the results below:
\begin{equation}
\nn
\tau_G(Y^{1,0}) = i ~, \tau_G(Y^{2,0}) = 2 i ~, \quad
\tau_G(Y^{3,0}) = \exp(\frac{2\pi i}{3}) ~, \tau_G(Y^{4,0}) = 2 i ~, \quad
\tau_G(Y^{2,2}) = \frac{2 i}{\sqrt{3}} ~.
\end{equation}
Comparing with the known results for $\tau_R$, we find exact agreement.
This should not surprise us given our success in the previous sections.
The spaces $Y^{p,0}$ are simply $\mZ_p$ orbifolds of the conifold, and $Y^{1,0}$ is just the conifold.
The space $Y^{p,p}$ are cones over lens spaces and are simply the quotient $\IC^3 / \mZ_{2p}$.
As orbifolds of the geometries that we have explicitly studied above, it is guaranteed that $\tau_G = \tau_R$.
At this point, it is interesting to recall that for the geometric counterpart of $a$-maximization the remainder function $g$ does not play any role and simply drops.
On the contrary, in our case we generically do explicitly need the full form of the symplectic potential --- \textit{i.e.}, including $g$.
More explicitly, while for the conifold and its orbifolds, $g$ plays no role, for the $Y^{p,p}$ spaces we find the correct result only upon considering the full $G$ including $g$.

What about something non-symmetric like $Y^{3,1}$?
We find that
\begin{equation}
\begin{array}{l}
\tau_G(Y^{3,1}) = -1+\frac{1}{3} i \sqrt{2+\sqrt{\frac{11}{3}}} ~, \\
\tau_R(Y^{3,1}) = -\frac{1}{2} i \left(1-2 \cos \left(\sqrt{\frac{11}{3}} \pi \right)+2 \cos \left(2
   \sqrt{\frac{11}{3}} \pi \right)+2 \cos \left(\sqrt{33} \pi \right)\right) \csc
   \left(\sqrt{33} \pi \right) ~.
\end{array}
\end{equation}
These are glaringly different expressions.
Can they be related to each other an $SL(2; \mZ)$ transformation which would mean that the tori are really the same?
In order to do so, we compute the Klein $j$-invariant for both cases.\footnote{Strictly speaking, we compute Klein's absolute invariant, without the conventional $1728$ prefactor so that $j(i) = 1$.}
We obtain that, numerically, $j(\tau_G(Y^{3,1})) \simeq 8.3796$ and
$j(\tau_R(Y^{3,1})) \simeq 8.4126$.
These are rather close real numbers, and given the complicated nature of the $j$-invariant, and the agreement in the other cases, this can not be a mere coincidence.
Let us for now bear this discrepancy in mind and accumulate more data.

\subsection{$L^{a,b,c}$, an extraordinary puzzle}
The next infinite family of affine Calabi--Yau manifolds well-known to the AdS/CFT community is the $L^{a,b,c}$ toric spaces \cite{bk, bfz}.
We again follow the nomenclature of \cite{Oota:2005mr}.
The symplectic potential can be computed starting with the outward pointing normal primitive vectors for $L^{a,b,c}$:
\begin{equation}
v_1=(1,\,1,\,0) ~, \quad
v_2=(1,\,a\,k,\,b) ~, \quad
v_3=(1,\,-\,a\,l,\,c) ~, \quad
v_4=(1,\,0,\,0) ~,
\end{equation}
where $k,\,l$ are integers such that $k\,c+b\,l=1$.
The Reeb vector $B$ is found by the minimization technique described in~\cite{Martelli:2005tp}; this is a vector $B = (3, b_2, b_3)$ which minimizes the functional
\begin{equation}
Z(b_2,b_3) = \frac{1}{24} \sum\limits_{i=1}^4
\frac{\det(\{w_{i-1}, w_i, w_{i+1}\})}{\det(\{B,w_{i-1},w_i\}) \det(\{B,w_i,w_{i+1}\})} ~,
\end{equation}
where $w_{1,2,3,4} = v_{1,2,3,4}$ and cyclically, $w_0 = v_4$, $w_5=v_1$, and where
the notation  $\det(\{a,\,b,\,c\})$ means the determinant of the matrix constructed by arranging $a,\,b,\,c$ as its rows.

The extremization will give some rather complicated quartics in $b_2$ and $b_3$, which we solve.
Finally, introducing
\begin{equation}
v_5=B-v_1-v_3 ~, \qquad
v_6=B-v_2-v_4 ~,
\end{equation}
for the extremized $B$ values, we obtain the full symplectic potential \eqref{Gpot}, including the summand $g$, in terms of coordinates $y=(y_1,\,y_2,\,y_3)$,
\begin{equation}
G=\frac{1}{2}\,\langle B,\,y\rangle\,\log\langle B,\,y\rangle +\frac{1}{2}\,\sum_{m=1}^3\,\langle v_{2\,m-1},\,y\rangle\,\log|\bar{x}-\bar{x}_m|+\frac{1}{2}\,\sum_{m=1}^3\,\langle v_{2\,m},\,y\rangle\,\log|\bar{y}-\bar{y}_m| ~,
\end{equation}
where $\langle\cdot,\,\cdot\rangle$ stands for the usual Cartesian product and
\begin{eqnarray}
&&
\bar{x}_1=-\frac{\det(\{v_1,\,v_5,\,v_6\})}{\det(\{v_3,\,v_5,\,v_6\})} ~, \qquad
\bar{x}_2=1 ~, \qquad
\bar{x}_3=-\frac{\det(\{v_1,\,v_5,\,v_6\})}{\det(\{v_1,\,v_3,\,v_6\})} ~, \\
&&
\bar{y}_1=1 ~, \qquad
\bar{y}_2=-1 ~, \qquad
\bar{y}_3=\frac{\beta+\alpha}{\beta-\alpha} ~,
\end{eqnarray}
where
\begin{equation}
\alpha=1+\frac{\det(\{v_2,\,v_3,\,v_4\})}{\det(\{v_3,\,v_4,\,v_6\})} ~, \qquad
\beta=1+\frac{\det(\{v_2,\,v_3,\,v_4\})}{\det(\{v_2,\,v_3,\,v_6\})} ~.
\end{equation}
Finally, $x$ and $y$ are implicitly defined by the equations
\begin{equation}
\langle v_2,\,y\rangle =\frac{\langle B,\,y\rangle}{2\,\alpha}\,(\alpha-\bar{x})\,(1-\bar{y}) ~, \qquad
\langle v_4,\,y\rangle =\frac{\langle B,\,y\rangle}{2\,\beta}\,(\beta-\bar{x})\,(1+\bar{y}) ~.
\end{equation}

Now, we have that $L^{a,a,a}=Y^{a,0}$, the orbifolds of the conifold
(the simplest case $L^{1,1,1}=Y^{1,0}$ is just the conifold, where we recovered above the expected result $\tau_G=i$).
Indeed, for the higher $a$ cases, we also have $\tau_G = \tau_R$.
For these cases, as described above, the remainder function $g$ does not play a r\^ole, and in fact with just $G^{{\rm can}}+G^{{\rm b}}$, we can reproduce this result.
However, moving to the next simplest case $Y^{p,p}$, the orbifolds of $\mathbb{C}^3$, we in fact have need the function $g$ to match $\tau_R$ and $\tau_G$.
Indeed, once we know the SLag fibration structure for a certain CY$_3$, its non-chiral orbifolds immediately follow.
By construction, these correspond to changing the period of the $\mathbb{T}^2$ coordinates.
Hence, we find that the complex structure of the relevant $\mathbb{T}^2$ follows the same pattern as in \cite{Hanany:2011ra}.
As remarked above, knowing that $\tau_G = \tau_R$ for $\mathbb{C}^3$ and the conifold, it is not a surprise that the same applies to their symmetric orbifolds.

Let us now move to a less symmetric and new example, namely that of $L^{1,2,1}$, otherwise known as the suspended pinched point (SPP).
This corresponds to D$3$-branes on the generalized conifold with the defining equation $xy=uv^2$.
Using the procedure above, we find that the geometric $\tau_G$ of SPP is
\begin{equation}
\tau_G=-\frac{1}{2}+\frac{i}{2}\,\sqrt{3\,(2+\sqrt{3})} \quad \Longrightarrow \quad j(\tau_G)\approx -20.8416 ~.
\end{equation}

Luckily the analytic expression for the complex structure of the isoradial dimer has been computed from field theory \cite{Jejjala:2010vb}.
It turns out that the $L^{a,b,a}$ subfamily has fields with R-charges~\cite{franco}
\be
R(u_1) = R(y) = \frac13 \frac{b-2a+w}{b-a} ~, \qquad
R(u_2) = R(z) = \frac12 R(v_1) = \frac13 \frac{2b-a-w}{b-a} ~,
\ee
where $w = \sqrt{a^2+b^2-ab}$.
Defining $\chi_i = \exp(i\frac{\pi}{2}R(u_i))$, we compute~\cite{Jejjala:2010vb}
\be
\tau_R = \frac{(\chi_1 + \chi_1^{-1})(1 + \chi_2^{-2})}{b(\chi_1 + \chi_1^{-1})\chi_2^{-2} - a(\chi_2 + \chi_2^{-1})\chi_1\chi_2 - (b-a)(\chi_2^2 + \chi_2^{-2})\chi_1} ~.
\ee
For $L^{1,2,1}$, we find
\begin{equation}
\tau_R(L^{1,2,1}) = \frac{1}{2} i \left(i+2 \tan \left(\frac{\pi }{2 \sqrt{3}}\right)+\cot \left(\frac{\pi
   }{2 \sqrt{3}}\right)\right) ~,
\end{equation}
and hence $j(\tau_R)\approx -20.3559$.
Thus, once more we find the remarkable proximity between $j(\tau_G)$ and $j(\tau_R)$.

We can at this point ask ourselves whether all this is due to numerical error.
After all, Klein's invariant $j$-function is a complicated non-linear function of its argument.
To settle this matter, we recall the discussion in Section~7 of~\cite{Jejjala:2010vb}.
Making use of the \textit{four exponentials conjecture}, it was argued that $\tau_R$ for SPP arising from the dimer construction is a transcendental number.
For reference, we leave a detailed account of this proof for the current case to Appendix~\ref{s:4exp}.
On the other hand, it is clear from our discussion above that whatever result we get from the geometrical construction, $\tau_G$ is going to be an algebraic number simply because $G$ only contains algebraic numbers and all subsequent manipulation to find $\tau_G$ involve only algebraic (in fact, quadratic) functions.
Now the $j$-invariant of two complex numbers are the same if and only if the two numbers are related by an $SL(2;\mZ)$ M\"obius transformation $\tau \mapsto \frac{a \tau + b}{c \tau + d}$ with $ad-bc=1$ and $a,b,c,d \in \mZ$.
Therefore, clearly, an algebraic number can not have the same $j$-invariant as a transcendental one.

Thus, assuming the four exponentials conjecture is true --- no counterexample to this conjecture is known to exist and it is widely accepted in the mathematical community --- $j(\tau_G)$ and $j(\tau_R)$ cannot match in this case, and we exclude numerical errors.
Nevertheless, it is truly remarkable that the geometry gives a torus so close to the field theory one.
Is there a fundamental explanation for this discrepancy?

In order to quantify the difference between $\tau_G$ and $\tau_R$, we focus on the $L^{1,b,1}$ family.
We can prove analytically here that
\begin{equation}\label{reJ}
\mbox{Re}(\tau_G) = \frac{b-1}{2} ~,
\end{equation}
while $\mbox{Im}(\tau_G)$ lives in some high even-degree extension of $\mQ$.
Nevertheless, \eqref{reJ} guarantees that $j$ will at least be real.\footnote{
To see this, we need merely look at the Laurent expansion of the $j$-function in terms of $q = e^{2\pi i\tau}$.
We see that $q$ and therefore $j(\tau)$ is real whenever the real part of $\tau$ is an integer or a half-integer.}
By a further modular transformation, we can always put $\mbox{Re}(\tau_G)$ to $0$ or $\frac12$.
In turn, from the field theory side, $\tau_R$ will be very complicated transcendental numbers.
While the real part of $\tau$ is of course not an $SL(2,{\mathbb Z})$ invariant and therefore not in itself a physically meaningful quantity, it may be convenient to set the real parts of $\tau_R$ and $\tau_G$ equal to each other via a modular transformation in order to facilitate a direct comparison of the two.
Quite surprisingly, the $SL(2,\,\mathbb{Z})$ transformation with $\{a=2,\,b=1,\,c=-1,\,d=0\}$ (plus repeated action with the $T$ generator) brings the real part to the form
\begin{equation}\label{reJR}
\mbox{Re}(\tau_R) = \frac{b-1}{2} ~.
\end{equation}
Hence, we are ensured that also the field theory $j$ will also be real.
Thus, we can, in the precise sense described above, associate the discrepancy between $\tau_R$ and $\tau_G$ to a small mismatch in their imaginary parts.

\begin{figure}[h!]
\centering
\includegraphics[scale=.7]{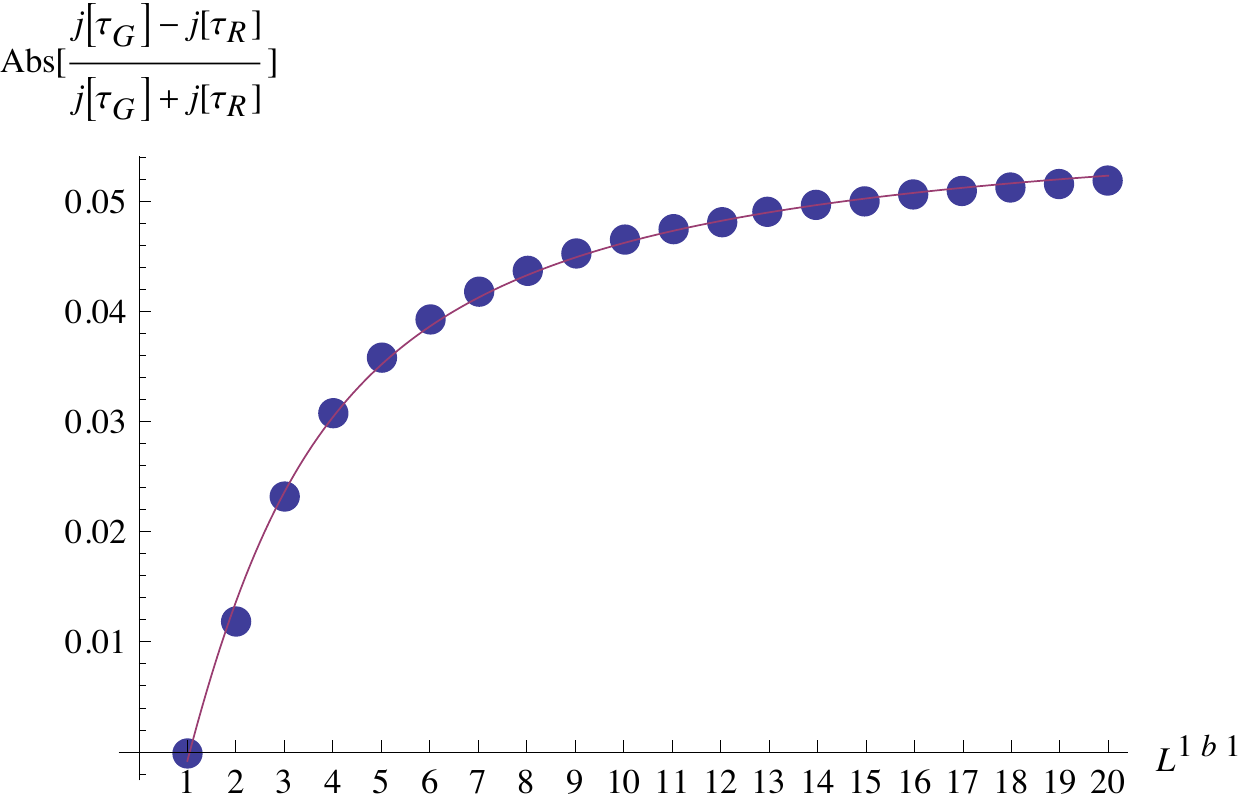}
\caption{For the spaces $L^{1,b,1}$, a measure of the difference between geometric torus, with complex structure $\tau_G$, and the QFT torus, with complex structure $\tau_R$.}
\label{DifferencesFit}
\end{figure}

We can compare the $j$-invariant of $\tau_G$ and $\tau_R$ for $L^{1,b,1}$ for various values of $b$.
In Figure~\ref{DifferencesFit}, we plot the quantity of the absolute value of the ratio of differences and find a beautiful fit to
\begin{equation}
\left| \frac{j(\tau_G) - j(\tau_R)}{j(\tau_G) + j(\tau_R)} \right|
\simeq -0.02+0.05\,\arctan\,0.4\,b ~.
\end{equation}

One interesting consequence is that the difference between $\tau_G$ and $\tau_R$, at least for the generalized conifold family under consideration, asymptotically saturates.

\section{Conclusions and prospects}
\label{sec:conclusions}

Dimer models have played a prominent role in the understanding of $\mathcal{N}=1$ SCFTs dual to D$3$-branes probing toric CY$_3$ singularities.
Their physical appearance was understood in~\cite{Feng:2005gw}.
In particular, the torus where the dimer lives was identified.
Our purpose in this letter is to take this identification further and metrically identify the torus.
Once the metric is known, it is natural to compute the complex structure, which might be expected to match the complex structure obtained from field theory considerations~\cite{Hanany:2011bs}.
Indeed, this is the very first step towards a full metric identification of the dimer from a geometrical perspective, hoping for a deeper understanding of some of its still somewhat mysterious properties, such as the nature of the isoradial embedding.

In \cite{Jejjala:2010vb} the combinatorial properties of dimers were first explored.
Since the dimer can be encoded in a very economical way in three permutations --- each capturing, respectively black vertices, white vertices, and faces --- through the Belyi theorem, it was argued that all the information can in fact be encoded in a Belyi pair consisting on a ``worldsheet'' torus together with a map from this ``worldsheet'' torus into a ``target'' $\mathbb{P}^1$ ramified only over $\{0,\,1,\,\infty\}$ such that the ramification data reproduces in a specific sense the combinatorial structure.
(For further details see \cite{Jejjala:2010vb}.) The beauty of the Belyi construction is that the existence of such map ensures that the worldsheet torus can be defined over $\overline{\mathbb{Q}}$.
Furthermore, this worldsheet torus is rigid, and so we naturally obtain, associated to each SCFT through ths torus serving as worldsheet for its Belyi map, another complex structure parameter $\tau_B$.
While in \cite{Jejjala:2010vb} it was speculated that $\tau_B$ might equal $\tau_R$, it was further shown in \cite{Hanany:2011ra} not to be true.
Thus, in view of our findings in this note, we have a triple of complex structures assigned to each SCFT, namely $\{\tau_R,\,\tau_G,\,\tau_B\}$.
We have seen that, in a precise sense, $\tau_R\sim \tau_G$.
It remains to fully clarify the nature of this agreement/disagreement as well the hypothetical connection to $\tau_B$.

In our case, a very simple reasoning led us to propose a certain $\mathbb{T}^2$ arising from the SLag fibration structure of the CY$_3$ as that where the dimer lives.
Indeed, for the simplest cases we found the torus expected from the field theory arguments.
Furthermore, we find that this torus behaves in the appropriate way under orbifolding thus reproducing the field theory patters for its complex structure.
However, moving to generic geometries we found this proposed torus to be almost but not exactly the expected one.
Indeed, focusing on the complex structure, we found that an ``$SL(2,\,\mathbb{Z})$ frame'' exists where both $\tau_R$ and $\tau_G$ have the same real part, while the imaginary part differs by a very small amount.
This is reflected in a more $SL(2,\,\mathbb{Z})$ invariant way in that the Klein $j$-invariants come to be surpassingly close to each other.
One must ask whether this discrepancy just signals that we have simply identified the torus incorrectly.
However, given the highly non-linear nature of the Klein invariant, or the tiny discrepancy in only the imaginary part of the complex structure, it would seem a cosmic coincidence to repeatedly have this almost matching purely by chance.

We note that the constructions in~\cite{Feng:2005gw} and~\cite{Imamura:2007dc} lie, respectively, three and two $T$-dualities away from the original IIB setup of D$3$-branes at the tip of the Calabi--Yau cone.
This raises the question as to which frame the $\mathbb{T}^2$ on which the dimer is drawn actually lives.
Na\"{\i}vely, two $T$-dualities would leave $\tau$ invariant up to an $SL(2,\,\mathbb{Z})$ transformation, thus suggesting that considering the original CY$_3$ should be enough.
In any case, this is the Calabi--Yau that is technically accessible in an explicit way.
Moreover, one would be naturally inclined to consider the original CY$_3$ as it is only in this frame that we have an AdS$_5$ space.
Recall that the complex structure of the $\mathbb{T}^2$ is fixed in field theory by the R-charges of the fields at the SCFT point, which on the other hand matches volumes in the IIB geometry.
However, we stress that one possible reason for the disagreement might simply be that we are looking to the wrong ``duality frame.''
At any rate, it is very surprising how close our ``wrong torus'' comes.
It is only slightly different in the imaginary part of the complex structure.
This demands an explanation. 

Uncovering the nature of the reason for the small disagreement would be extremely interesting.
It seems one fundamental problem we face is that it is hard to find a ``microscopic quantification'' of the disagreement.
In other words, should we consider the Klein invariant as we have done in the paper?
Or should we rather consider some other modular invariant?
In fact, motivated by the fact that the $\tau_R$ comes in terms of transcendental numbers of the form $e^{i\,\frac{\pi}{2\,\sqrt{3}}}$, one natural place to look for such corrections might be in terms of instanton contributions.
These would arise upon resolving the singularity at the tip of the cone to introduce K\"ahler moduli.
However, it is unclear what exactly those instantons would contribute and what their exact nature is.
We leave this very interesting problem open.

\section*{Acknowledgements}

We wish to thank James Sparks for being generous with his time and offering important insights into the geometric analysis.
We are grateful to Amihay Hanany, Jurgis Pasukonis, and Sanjaye Ramgoolam for collaboration on related topics.
YHH would like to thank the
Science and Technology Facilities Council, UK, for an Advanced Fellowship and grant ST/J00037X/1, the Chinese Ministry of Education, for a Chang-Jiang Chair Professorship at NanKai University, as well as City University, London and Merton College, Oxford, for their enduring support.
YHH and VJ jointly acknowledge NSF grant CCF-1048082.
The work of VJ is based upon research supported by the South African Research Chairs Initiative of the Department of Science and Technology and National Research Foundation.
VJ as well thanks Queen Mary, University of London and STFC grant ST/G000565/1 for supporting his work during the early stages of this project.
DRG is supported by the Aly Kaufman fellowship. He also acknowledges partial support from the Israel Science Foundation through grant 392/09 and from the Spanish Ministry of Science through the research grant FPA2009-07122 and Spanish Consolider-Ingenio 2010 Programme CPAN (CSD2007-00042).

\begin{appendix}

\section{Complex structures of tilted tori}
\label{sec:sexytau}

Let us consider a generic $\mathbb{T}^2$.
The most general form for its metric is
\begin{equation}
ds^2=A\,d\phi_1^2+B\,d\phi_2^2+C\,d\phi_1\,d\phi_2 ~.
\end{equation}
Massaging the expression converts this into
\begin{equation}
ds^2=\frac{4\,A\,B-C^2}{4\,B}\,d\phi_1^2+B\,\Big( d\phi_2+\frac{C}{2\,B}\,d\phi_1 \Big)^2 ~.
\end{equation}

It is now convenient to introduce
\begin{equation}
\psi_1=\frac{\sqrt{4\,A\,B-C^2}}{2\,\sqrt{B}}\,\phi_1 ~, \qquad
\psi_2=\sqrt{B}\,\phi_2
\end{equation}
whose identifications are $\psi_i\sim \psi_i+\Delta_i$ with
\begin{equation}
\Delta_1=\frac{\sqrt{4\,A\,B-C^2}}{2\,\sqrt{B}}\,2\,\pi ~, \qquad
\Delta_2=\sqrt{B}\,2\,\pi ~.
\end{equation}
In these coordinates,
\begin{equation}
ds^2=d\psi_1^2+\Big(d\psi_2+\gamma\,\,d\psi_1\Big)^2 ~, \qquad
\gamma=\frac{C}{\sqrt{4\,A\,B-C^2}} ~.
\end{equation}

This corresponds to a torus generated by
\begin{equation}
\vec{\ell}_1=\Delta_1\,(1,\,-\gamma) ~, \qquad
\vec{\ell}_2= \Delta_2\,(0,\,1) ~.
\end{equation}
Hence,
\begin{equation}
\tau=\frac{\Delta_1\,\gamma}{\Delta_2}\,(1+i\,\gamma^{-1}) ~.
\end{equation}
In terms of $A,\,B,\,C$ we find
\begin{equation}
\label{tau}
\tau=\frac{C}{2\,B}\,\Big(1+i\,\frac{\sqrt{4\,A\,B-C^2}}{C}\Big) ~.
\end{equation}

\section{The four exponentials conjecture}
\label{s:4exp}

The \textit{four exponentials conjecture} is one of the key consequences of Schanuel's conjecture and would constitute one of the most important results in number theory.
It states that given two pairs of complex numbers $(x_1, x_2)$ and $(y_1, y_2)$ such that each pair is linearly independent over $\mathbb{Q}$, then at least one of the numbers in the list
\begin{equation}
\{ e^{x_1 y_1} ~, e^{x_1 y_2} ~, e^{x_2 y_1} ~, e^{x_2 y_2} \} ~,
\end{equation}
is transcendental.

Now, consider our case of $L^{1,2,1}$.
We have that
\begin{equation}
\tau_R = \frac{1}{2} i \left(i+2 \tan \left(\frac{\pi }{2 \sqrt{3}}\right)+\cot \left(\frac{\pi}{2 \sqrt{3}}\right)\right) = \frac{1-3x^2}{x^4-1} ~,
\label{eq:taur4}
\end{equation}
where $x = \exp( i \frac{\pi}{2 \sqrt{3}})$.
Noting that this is an algebraic (in fact, rational) function in the single complex number $x$, it suffices to show that $x$ is transcendental to imply that $\tau_R$ is also.

Let $(x_1, x_2) = (1, \frac{1}{2 \sqrt{3}})$ and $(y_1,y_2) = (\pi i , \frac{\pi i}{2 \sqrt{3}})$, we form the list of three number ($x$ appears twice):
\be
e^{\pi i} = -1 ~,
e^{i \frac{\pi}{2 \sqrt{3}}} = x ~,
e^{\frac{\pi i}{12}} =  \frac{\sqrt3+1}{2 \sqrt{2}}+i \left(\frac{\sqrt3-1}{2 \sqrt{2}}\right) ~.
\ee
Now, the conjecture states that at least one of these must be transcendental and seeing the first and last to be clearly algebraic, $x$ must thus be the transcendental one.
Rewriting~\eref{eq:taur4}, we have:
\be
\tau_R \, x^4 + 3x^2 - ( \tau_R + 1) = 0 ~.
\ee
If $\tau_R$ were algebraic, then $x$ must be algebraic since $\overline{\mathbb Q}$ is algebraically closed.
Because $x$ is not algebraic assuming the four exponentials conjecture, the contrapositive applies and $\tau_R$ must be transcendental.

\end{appendix}

\end{document}